\documentclass[12pt]{article}

\usepackage{amsmath,amssymb,graphicx} 
\usepackage{epsf}
\usepackage{pstricks}
\usepackage{cite}

\newcommand{\beq}{\begin{eqnarray}}
\newcommand{\eeq}{\end{eqnarray}}

\newcommand{\centeron}[2]{{\setbox0=\hbox{#1}\setbox1=\hbox{#2}\ifdim
                              \wd1>\wd0\kern.5\wd1\kern-.5\wd0\fi \copy0

\kern-.5\wd0\kern-.5\wd1\copy1\ifdim\wd0>\wd1
                              \kern.5\wd0\kern-.5\wd1\fi}}
\newcommand{\ltap}{\>\centeron{\raise.35ex\hbox{$<$}}
                      {\lower.65ex\hbox{$\sim$}}\>}
\newcommand{\gtap}{\>\centeron{\raise.35ex\hbox{$>$}}
                      {\lower.65ex\hbox{$\sim$}}\>}

\newcommand\ZZ{\hbox{\zfont Z\kern-.4emZ}}
\font\zfont = cmss10 
\newcommand{\sfrac}[2]{{\textstyle\frac{#1}{#2}}}

\def\tv#1{\vrule height #1pt depth 5pt width 0pt}

\begin{document}
\begin{titlepage}
\begin{flushright}
{\tt hep-ph/0401160}\\
MCTP-04-03 \\
Saclay t04/009\\
\end{flushright}

\vskip.5cm
\begin{center}
{\huge \bf  Oblique Corrections from Higgsless Models in Warped Space}
\vskip.1cm
\end{center}
\vskip0.2cm

\begin{center}
{\bf
{Giacomo Cacciapaglia}$^{a}$, {Csaba Cs\'aki}$^{a}$,
{Christophe Grojean}$^{b,c}$,\\
{\rm and}
John Terning$^{d}$}
\end{center}
\vskip 8pt

\begin{center}
$^{a}$ {\it Institute for High Energy Phenomenology\\
Newman Laboratory of Elementary Particle Physics\\
Cornell University, Ithaca, NY 14853, USA } \\
\vspace*{0.1cm}
$^{b}$ {\it Service de Physique Th\'eorique,
CEA Saclay, F91191 Gif--sur--Yvette, France} \\
\vspace*{0.1cm}
$^{c}$ {\it Michigan Center for Theoretical Physics,
Ann Arbor, MI 48109, USA}\\
\vspace*{0.1cm}
$^{d}$ {\it Theory Division T-8, Los Alamos National Laboratory, Los
Alamos,
NM 87545, USA} \\
\vspace*{0.3cm}
{\tt  cacciapa@lepp.cornell.edu, csaki@lepp.cornell.edu,
grojean@spht.saclay.cea.fr, terning@lanl.gov}
\end{center}

\vglue 0.3truecm

\begin{abstract}
\vskip 3pt
\noindent
We calculate the tree-level oblique corrections to electroweak 
precision observables generated in higgless models of
electroweak symmetry breaking with a 5D SU(2)$_L\times$SU(2)$_R$ 
$\times$U(1)$_{B-L}$ gauge group on a warped
background. In the absence of brane induced kinetic terms (and equal 
left and right gauge couplings) we find the
$S$ parameter to be $\sim 1.15$, while $T \sim U\sim 0$, as in 
technicolor theories. Planck brane induced kinetic terms
and unequal left-right couplings can lower $S$, however for 
sufficiently low values of $S$ tree-level unitarity
will be lost. A kinetic term localized on the TeV brane for SU(2)$_D$ 
will generically increase $S$, however
an induced kinetic term for U(1)$_{B-L}$ on the TeV brane will lower 
$S$. With an
appropriate choice of the value of this
induced kinetic term $S\sim 0$ can be achieved. In this case the mass 
of the lowest Z$'$ mode will be lowered to about~$300$~GeV.

\end{abstract}

\end{titlepage}

\newpage


\section{Introduction}
\label{sec:intro}
\setcounter{equation}{0}
\setcounter{footnote}{0}

Finding the correct mechanism for electroweak symmetry breaking is 
perhaps the most important problem
facing particle physics. The simplest possibility is via a standard 
model (SM) Higgs field,
however the mass of such a Higgs is unstable to radiative corrections. 
The usual ways to overcome this
problem is either to assume that the Higgs mass is stabilized by 
supersymmetry, or that electroweak symmetry
is broken dynamically by some interaction becoming strong around the 
TeV scale. Extra dimensions may also offer possible ways
of stabilizing the Higgs mass against radiative corrections, either by 
lowering the scale of gravity in large extra
dimensions~\cite{ADD}, by localizing gravity away from the SM fields in 
the RS model~\cite{RS}, or by the Higgs secretly being the scalar
component $A_5$ of an extra dimensional gauge field~\cite{A5Higgs}
(the 4D implementation of this last approach leads to the recently
proposed~\cite{littleHiggs} little Higgs models). However, extra 
dimensions could not only offer new ways to stabilize the Higgs mass, 
but also a
completely new mechanism to break the symmetry itself without the 
existence of a Higgs: boundary conditions
at the endpoints of a finite extra dimension could break a gauge 
symmetry, in which case unitarity of the scattering
amplitudes of massive gauge bosons (GB's) is not cured by the exchange 
of the physical Higgs
scalar, but rather by the exchange of a tower of massive KK gauge 
bosons~\cite{CGMPT} (see also \cite{otherunitarity}).
A toy model with massive W and Z bosons of this
sort was presented in~\cite{CGMPT}. In order to  find a realistic 
higgsless model of electroweak symmetry breaking one needs
to overcome several problems. The first is the question of how to 
ensure that the boundary conditions (BC's)
give the right $M_W/M_Z$ mass ratio. The correct tree-level prediction 
is usually ensured by a custodial SU(2) global
symmetry in the sector that breaks the electroweak symmetry. The way to 
implement custodial SU(2) in an extra dimensional
theory is by putting SU(2)$_L\times$SU(2)$_R$ gauge bosons in an 
anti-de Sitter (AdS) background~\cite{ADMS}. The reason
behind this lies in the holographic interpretation of an AdS bulk: it 
is supposed to correspond to a four dimensional
conformal field theory (CFT) which has a global symmetry dictated by 
the gauge symmetries in the bulk. Those symmetries
broken on the Planck brane (the UV brane) will be interpreted as global 
symmetries, while those unbroken will remain as weakly
gauged global symmetries of the CFT all the way down to the TeV scale. 
Based on this insight, \cite{CGPT} considered
a 5D gauge theory on an AdS$_5$ background with 
SU(2)$_L\times$SU(2)$_R\times$U(1)$_{B-L}$ gauge bosons in the bulk.
SU(2)$_R\times$U(1)$_{B-L}$ is broken to U(1)$_Y$ on the Planck brane, 
leaving the broken generators as global
symmetries (and thus ensuring the custodial SU(2) symmetry necessary 
for recovering the GB mass ratios). On the TeV brane
SU(2)$_L\times$SU(2)$_R$ is broken to SU(2)$_D$ thereby breaking 
electroweak symmetry.\footnote{Alternatively a model in flat space can 
be constructed, where the approximate custodial symmetry is 
enforced~\cite{BPR}
by introducing large kinetic terms localized on the 
SU(2)$_L\times$U(1)$_Y$ brane with the effect of pushing the 
wavefunctions away from the location where the custodial symmetry is 
broken.} This theory could be considered
as the AdS/CFT dual of walking technicolor \cite{walking}: the CFT runs 
slowly until energies where the TeV brane appears and
electroweak symmetry is broken. It was shown in~\cite{CGPT} (and 
in~\cite{Nomura} for the same model with an enlarged
parameter space) that the leading SM predictions are recovered in this 
setup: one gets the correct W/Z mass ratio and
coupling to fermions localized close to the Planck brane. The KK modes 
of the W and Z are heavy enough to have escaped
direct detection, but just light enough to unitarize the scattering 
amplitudes. For the simplest choice of parameters
the first of these KK modes would appear at around 1.2 TeV.

Two important issues have to be addressed to make the model completely 
realistic: how to generate fermion masses in the
absence of a Higgs scalar, and how large the corrections to electroweak 
precision observables would be. In~\cite{CGHST} (and also
in~\cite{CGPT,BPR,Nomura})
it has been shown that it is possible to obtain a realistic fermion 
mass spectrum in this model: the
unbroken gauge group on the TeV brane is non-chiral, so an explicit 
mass term can be added there. In order to lift the
SU(2)$_D$ degeneracy, mixing with Planck brane localized fermions can 
be added (which is equivalent to adding different
brane localized kinetic terms for the right handed up vs. down type 
quarks, or neutrinos vs. charged leptons). The issue of
electroweak precision observables in higgsless theories was first 
discussed in~\cite{BPR}, where it was argued that
in generic higgsless models there will be an $S$ parameter of order one 
generated (analogously to technicolor theories). In this
paper we investigate the tree-level oblique corrections to electroweak 
precision observables in the warped higgsless model in terms
  of the  input parameters of the theory. We find, that in the absence 
of brane induced kinetic terms (and equal left and right
couplings) $S\sim 1.15, T\sim U\sim 0$. Unequal L,R couplings or Planck 
brane induced kinetic terms can lower $S$ to acceptable
values only at the price of losing tree-level unitarity (thus entering 
a strong coupling regime). TeV brane induced SU(2)
kinetic terms generically raise the value of $S$, however a TeV brane 
induced kinetic term for the U(1)$_{B-L}$ gauge group
lowers $S$. By choosing an appropriate value for this coupling one can 
obtain $S\sim 0$ (and a moderately negative $T$).
In this case the mass of the lightest Z$'$ KK mode is lowered to about 
300~GeV.

Note however, that while this analysis does give a large fraction of 
the corrections to electroweak precision observables, one
should not directly compare the values of $S,T,U$ obtained here to the 
usually quoted experimental values. The reason is that
the experimentally allowed values of $S,T,U$ are usually extracted from 
the data assuming the existence of a SM Higgs with
a certain mass (usually 115~GeV). Here however there is no Higgs 
particle, so its contribution should be subtracted
from the analysis. This procedure is however complicated by the fact 
that only the sum of the gauge plus Higgs sector gives a
finite and gauge invariant contribution to the oblique parameters, so 
the right approach would be to replace this sum with
the 1-loop contribution of the full KK tower of the W and Z. In 
addition, depending on the actual fermion mass model, there
can be large non-oblique corrections to the interactions of the third 
generation, in particular to the $Zb\bar{b}$ vertex.
These issues are left for further investigation. Further interesting 
questions regarding higgsless models have been
discussed in~\cite{BRST,spurion,MSU}: non-perturbative arguments for 
unitarity of models with BC breaking
of gauge symmetries have been presented in~\cite{BRST,spurion}, while 
deconstruction of the higgsless models was
considered in~\cite{MSU} (see also~\cite{CGMPT}).

This paper is organized as follows: in Section 2 we discuss the setup 
and the expansion of the wave functions to sub-leading
order in $ 1/\log (R'/R)\sim 1/30$. In section 3 we present the 
calculation of the oblique corrections for the simplest
point on the parameter space. In Section 4 we extend this calculation 
to asymmetric gauge couplings, and more
importantly including brane induced kinetic terms. We show how $S$ can 
be lowered using the U(1)$_{B-L}$ induced kinetic term on the
TeV brane, and briefly discuss the properties of the lightest Z$'$ mode 
for the case with small $S$. We conclude in Section 5.

While this paper was in preparation, two related studies~\cite{DHLR,BN}
  of the oblique corrections to electroweak precision observables in 
higgsless models appeared, which have a
considerable overlap with our work.

\section{The model}
\setcounter{equation}{0}
\setcounter{footnote}{0}
We will consider the model discussed in
\cite{CGPT}: an SU(2)$_L\times$SU(2)$_R\times$U(1)$_{B-L}$ gauge theory 
on a fixed AdS$_5$ background.
We will use the conformally flat metric
\beq
ds^2=  \left( \frac{R}{z}\right)^2   \Big( \eta_{\mu \nu} dx^\mu dx^\nu 
- dz^2 \Big)
\eeq
where $z$ is on the interval $[R,R^\prime]$. The bulk Lagrangian for 
the gauge fields
(after decomposing the 5D gauge boson into a 4D gauge
boson $A_\mu^a$ and a 4D scalar $A_5^a$ in the adjoint representation, 
and adding the appropriate
gauge fixing terms in R$_\xi$ gauge) are of the form
\begin{equation}
           \label{eq:lagr}
\mathcal{S}=
\int d^4x\,\int_R^{R^\prime} dz\,  \frac{R}{z} \, \left(
-\frac{1}{4} F_{\mu\nu}^a F^{a\mu\nu} -\frac{1}{2} F_{5\nu}^a F^{a5\nu}-
\frac{1}{2\xi} (\partial_\mu A^{a\mu} -\xi \partial_5 A_5^a)^2 \right),
\end{equation}
where $M=0,1,2,3,5$ denotes 5D Lorentz indices while $\mu=0,1,2,3$ are 
used to denote usual 4D Lorentz indices. $F^a_{MN}=\partial_M A_N^a -  
\partial_N A_M^a
+g_5 f^{abc}\, A^b_M A^c_N$, and the $f^{abc}$'s
are the structure constants of the gauge group, and we have 
$F^a_{MN}=(W^{L,a}_{MN},W^{R,a}_{MN},B_{MN})$
for the SU(2)$_L$, SU(2)$_R$, U(1)$_{B-L}$ gauge groups. We will denote 
the 5D gauge couplings of these groups by
$g_{5L},g_{5R}$ and $\tilde{g}_5$.
Taking $\xi\to \infty$ will result in the unitary gauge, where all the
KK modes of the scalars fields $A^a_5$ are unphysical (they become the
longitudinal modes of the 4D gauge bosons), except if there is a zero 
mode for the
$A_5$'s. A zero mode in $A_5$ would correspond to a Goldstone boson in 
the
holographic interpretation. Such Goldstone modes exist if some of the 
bulk
gauge symmetries are broken both on the Planck brane and the TeV brane. 
In
the model considered below this will not be the case,
and so will assume that every $A_5^a$ mode is massive, and thus
that all the $A_5$'s are eliminated in the unitary gauge.

We denote by
$A^{R\,a}_{M}$, $A^{L\,a}_{M}$ and $B_M$ the gauge bosons of
$SU(2)_R$, $SU(2)_L$ and $U(1)_{B-L}$ respectively. For simplicity we 
will first assume
$g_{5L}=g_{5R}=g_5$ (later on we will relax this assumption).
We impose the following BC's:

\begin{eqnarray}
&
{\rm at }\  z=R^\prime:
&
\left\{
\begin{array}{l}
\partial_z (A^{L\,a}_\mu +A^{R\,a}_\mu)= 0, \
A^{L\,a}_\mu  - A^{R\,a}_\mu =0, \
\partial_z B_\mu = 0,
\\
\tv{15}
(A^{L\,a}_5 +A^{R\,a}_5) = 0, \
\partial_z (A^{L\,a}_5 -A^{R\,a}_5)  = 0, \
B_5 = 0.
\end{array}
\right.
\label{bc1}\\
&
{\rm at }\ z=R:
&
\left\{
\begin{array}{l}
\partial_5 A^{L\,a}_\mu=0, \
       A^{R\,1,2}_\mu=0,
\\
\tv{15}
\partial_z (g_5  B_\mu + \tilde g_5  A^{R\,3}_\mu )= 0, \
\tilde g_5 B_\mu - g_5 A^{R\,3}_\mu=0,
\\
\tv{15}
A^{L\,a}_5=0, \ A^{R\,a}_5=0, \ B_5 = 0.
\end{array}
\right.
\label{bc2}
\end{eqnarray}

The KK expansion is given by
\begin{eqnarray}
             \label{eq:KKB}
B_\mu (x,z) & = & \frac{1}{\tilde g_5}\,  a_0 \gamma_\mu (x)
+   \sum_{k=1}^{\infty} \psi^{(B)}_k (z)  \, Z^{(k)}_\mu (x)\, ,
\\
             \label{eq:KKAL3}
A^{L\, 3}_\mu (x,z) & = &
\frac{1}{g_5} \, a_0 \gamma_\mu (x)
+   \sum_{k=1}^{\infty} \psi^{(L3)}_k (z) \, Z^{(k)}_\mu (x) \, ,
\\
             \label{eq:KKAR3}
A^{R\, 3}_\mu (x,z) & = &
\frac{1}{g_5}  \, a_0 \gamma_\mu (x)
+    \sum_{k=1}^{\infty}  \psi^{(R3)}_k (z) \, Z^{(k)}_\mu (x) \, ,
\\
             \label{eq:KKALpm}
A^{L\, \pm}_\mu (x,z) & = &
    \sum_{k=1}^{\infty}  \psi^{(L\pm)}_k (z) \, W^{(k)\, \pm}_\mu (x) \, 
,
\\
             \label{eq:KKARpm}
A^{R\, \pm}_\mu (x,z) & = &
    \sum_{k=1}^{\infty} \psi^{(R\pm)}_k  (z) \, W^{(k)\, \pm}_\mu (x)  \,
.
\end{eqnarray}

The Euclidean bulk equation of motion satisfied
by spin-1 fields in AdS space is
\begin{equation}
(\partial_z^2 - \frac{1}{z} \partial_z  +q^2) \psi(z)=0,
\end{equation}
where the solutions in the bulk are assumed to be of the form
$A^a_\mu (q) e^{-iqx} \psi(z)$. The  KK mode expansion is given by
the solutions to this equation which are
of the form
\beq
	\label{eq:Bwv}
\psi^{(A)}_k(z)=z\left(a^{(A)}_k J_1(q_k z)+b^{(A)}_k Y_1(q_k
z)\right)~,
\eeq
where $A$ labels the corresponding gauge boson.

Next we want to do an expansion in $ 1/\log (R'/R)$ for the wave 
functions of the
gauge bosons. When $R'\to \infty$, the TeV brane is sent to
infinity, and thus there is no electroweak symmetry breaking. So in 
this limit
$M_W,M_Z\to 0$, and the wave functions will be exactly flat.
For finite $R'$, $M_W^2, M_Z^2 \propto  1/\log (R'/R)$ \cite{CGPT}. 
This means
that the argument of the gauge boson wave function $M z$ is at most of 
order
$ 1/\log (R'/R)$, thus an expansion in small arguments for the Bessel 
functions is
justified and all quantities related to the light gauge bosons should 
have a
well-defined expansion in $ 1/\log (R'/R)$. In \cite{CGPT} we have 
shown that the
leading terms $\sim  1/\log (R'/R)$ in this log expansion exactly 
reproduce the
SM mass relations and the SM couplings. The $1/\log^2 (R'/R)$ will then
correspond to the leading corrections to the electroweak precision 
observables.
In the rest of this section we will find the corrections to the gauge 
boson wave
functions to order $1/\log^2 (R'/R)$, which we will then use in the 
next section
to get the oblique correction parameters $S,T,U$ to electroweak 
precision
observables.

    From the expansion
for small arguments of the Bessel functions appearing
in~(\ref{eq:Bwv}), the
wavefunction of a mode with mass
$M  \ll 1/R^\prime$ can be written as (analogously to the expansions 
in~\cite{CGPT,CET}):
\beq
\psi^{(A)}(z) &\approx  &
c_0^{(A)}
+ M_A^2 z^2 \left( c_1^{(A)}   - \frac{c_0^{(A)} }{2}  \log(z/R)\right)
\nonumber\\
&&+M_A^4 z^4 \left( -\frac{c_1^{(A)}}{8}-\frac{3 c_0^{(A)}}{64}   +
\frac{c_0^{(A)} }{16}  \log(z/R)\right)
+ {\cal O} (M_A^6 z^6),
\eeq
with
$c_1^{(A)}$ at most ${\cal O} (c_0^{(A)}  \log(R^\prime/R))$, and $(M 
R^\prime)^2
\sim {\cal O}
(1/ \log(R^\prime/R))$. The integrals of these wave functions that will 
be
relevant for electroweak precision observables are given by
(suppressing the index $A$):
\beq
\int_R^{R^\prime}  \,  \frac{dz}{z} \psi(z)^2 &\approx&
c_0^{2} \log \left( \frac{R^\prime}{R}\right)+ M^2 R^{\prime 2} \left(
c_0 c_1
-\frac{1}{2}  c_0^2
    \log \sfrac{R^\prime}{R} \right) \\
&& +M^4 R^{\prime 4} \left( \frac{c_0^2 \left( \log
\sfrac{R^\prime}{R}\right)^2}{16}
-\frac{c_0 c_1 \log  \sfrac{R^\prime}{R}}{4}
+\frac{c_1^2}{4}   \right)
     + {\cal O} \left(\frac{1}{
\log \sfrac{R^\prime}{R}}\right)\nonumber\\
\int_R^{R^\prime}  \,  \frac{dz}{z} \left[\partial_z \psi(z)\right]^2
&\approx&
\left( \frac{ M^4 R^{\prime 2} }{2}-\frac{M^6 R^{\prime 4}}{8}  \right)
\left(
c_0^{2}  \left(\log \sfrac{R^\prime}{R} \right)^2
- 4 c_0 c_1
   \log \sfrac{R^\prime}{R}
+4 c_1^2  \right)\\
     &&+ {\cal O} \left(\frac{1}{
\left( \log \sfrac{R^\prime}{R}\right)^2 }
\right)\nonumber
\eeq

The boundary conditions on the bulk gauge fields give the following
results for the leading and next-to-leading log terms in the
wavefunction for the lightest charged
gauge bosons
\beq
    c_0^{(L\pm)} &\equiv& c_\pm~, \ \
    c_0^{(R\pm)} = {\cal O}\left(c_\pm \frac{R^2}{R^{\prime 2}}\right)
\approx  0~,
\label{eq:c0Lpm}\\
    c_1^{(L\pm)} &\approx& \frac{c_\pm}{4}~, \ \
    c_1^{(R\pm)} \approx  \frac{c_\pm}{2}
\log\left(\frac{R^\prime}{R}\right)
    +{\cal O}\left(\frac{c_\pm}{ \log\left(\frac{R^\prime}{R}\right)
}\right)  ~,
\eeq
while for the neutral gauge bosons we find in the same approximation
\beq
c_0^{(L3)} \equiv \, c~, \ \
c_0^{(R3)} \approx  -c  \, \frac{\tilde g_5^2}{g_5^2+ \tilde g_5^2}~, \
\
c_0^{(B)} \approx -c  \, \frac{g_5 \, \tilde g_5}{g_5^2+ \tilde g_5^2}~.
\eeq
To leading log order we also have:
\beq
c_1^{(L3)} & \approx & \, \frac{c}{4}~, \\
c_1^{(R3)} &\approx&\frac{ c}{2}  \, \frac{g_5^2 }{g_5^2+ \tilde
g_5^2}\log \frac{R^\prime}{R}-\frac{c}{4}  \, \frac{\tilde
g_5^2}{g_5^2+ \tilde
g_5^2}+\ldots ~, \\
c_1^{(B)} &\approx& -\frac{ c}{2}   \, \frac{g_5 \, \tilde g_5}{g_5^2+
\tilde
g_5^2} \log \frac{R^\prime}{R}-\frac{c}{4}   \, \frac{g_5 \, \tilde
g_5}{g_5^2+ \tilde
g_5^2} +\ldots  ~.
\label{eq:c1B}
\eeq

To leading order in $1/R$ and to next-to-leading order in
$1/\log \left(R^\prime/R \right)$, the lightest solution for this
equation for the mass of the $W^\pm$'s is
\begin{equation}
M_W^2 = \frac{1}{R^{\prime 2} \log \left(\frac{R^\prime}{R}\right)}
\left(1+\frac{3}{8\log\frac{R'}{R}} \right)\, ,
\end{equation}
while the lowest mass of the $Z$ tower is approximately given by
\begin{equation}
M_Z^2  = \frac{g_5^2+2 \tilde g_5^{2}}{g_5^2+ \tilde g_5^{2}}
\frac{1}{R^{\prime 2} \log \left(\frac{R^\prime}{R}\right)}
\left(1+\frac{3}{8\log\frac{R'}{R}}  \frac{g_5^2+2 \tilde
g_5^{2}}{g_5^2+ \tilde g_5^{2}}\right) \, .
\end{equation}
%

\section{Oblique Corrections to Precision Electroweak Observables}
\setcounter{equation}{0}
\setcounter{footnote}{0}

To calculate precision electroweak corrections we first need to choose
a ``renormalization convention." Our aim is to find a scheme in which 
all
corrections will be oblique (that is the effective 4D Lagrangian 
describing the coupled gauge boson-fermion system
only gets corrections in the gauge boson sector, but not the fermion 
sector). With bulk gauge kinetic terms
normalized by $1/g_5^2$ a very simple convention is to set the gauge
boson wavefunction to be one at the location of the fermion \cite{CET}.
Here we have chosen canonical normalization. What we need to require is 
that the couplings of Planck brane localized
fermions give exactly the leading order relations for the couplings. 
This definition will make the corrections oblique.
The couplings of an SU(2)$_L$ doublet fermion to the vector gauge bosons
are read off from the bulk covariant derivative evaluated at the 
location of the fermion, i.e., at the Planck brane:
\begin{multline}
\left(
g_5 T_3 A^{L3}_\mu + g_5 T_{\pm} A^{L \mp}_\mu + \frac{Y}{2}
\tilde g_5 B_\mu \right)_{|z=R} =\\
a_0 Q \gamma_\mu + g_5 \psi_1^{L\mp} (R) T_{\pm} W^{\mp}_\mu + g_5 
\psi_1^{(L3)} (R) \left( T_3 + \frac{\tilde g_5 \psi_1^{(B)} (R)}{g_5 
\psi_1^{(L3)} (R)} \frac{Y}{2} \right) Z_\mu~,
\label{eq:couplings}
\end{multline}
where we have used the KK expansion (\ref{eq:KKB})-(\ref{eq:KKARpm}) as 
well as the relation
$\tilde g_5 B_\mu  = g_5 A_\mu^{L3}$ for the photon wavefunction.
$Y$ is the SM hypercharge of the fermion, $Q=Y/2+T_{3L}$ its electric 
charge. Using the BC
$\tilde g_5 B_\mu  = g_5 A_\mu^{R3}$ on the Planck brane, an identical 
expression also holds
for the SU(2)$_L$ singlet fermions that are embedded into SU(2)$_R$ 
doublet representations (see Ref.~\cite{CGHST} for details).

We now have to compare these couplings with the SM ones. The first 
point to note is that
in this formula, the only quantity completely fixed by the BC's, and 
then independent on the overall normalizations of the W and the Z, is 
the ratio of the $Z$ couplings to the $T_3$ and $Y$ component of the 
fermion.
Thus, in order to recover the SM couplings for the fermions and ensure 
that besides the oblique corrections there are no
other corrections we will have to impose that the one ratio independent 
of the normalizations reproduces the SM result:
\beq \label{tgdef}
\frac{g'^2}{g^2} = - \frac{\tilde g_5 \psi_1^{(B)} (R)}{g_5 
\psi_1^{(L3)} (R)}~.
\eeq
Moreover, due to the unbroken U(1)$_{em}$ it is always possible to 
canonically normalize the photon kinetic term in the effective
4D Lagrangian, thus we choose
\beq
Z_\gamma&=&\int_R^{R'}[\Psi^{\gamma}]^2\left(\frac{R}{z}\right)
dz=\int_R^{R'}([a_0/{\tilde g_5}]^2+2[a_0/g_5]^2)
\left(\frac{R}{z}\right)  dz =1~.
\eeq
Since the photon wave function is flat it is easy to find its coupling 
to the Planck brane fermions
and thus define the electric charge ($1/e^2=1/g^2+1/g'^2$) as:
\beq \label{edef}
\frac{1}{e^2} = \frac{1}{a_0^2}=  \left( 
\frac{1}{\tilde{g}_5^2}+\frac{2}{g_5^2}
\right) R \log R'/R~.
\eeq
The two equations~(\ref{tgdef}) and (\ref{edef}) completely define the 
4 dimensional SM gauge couplings in terms of the 5D parameters. At the 
leading order in the log expansion, we recover
the relations obtained in~\cite{CGPT}:
\beq
g^2&=&
\frac{g_5^2}{R \log(R^\prime/R)},
\label{SM1}\\
e^2&=& \frac{g_5^2 \tilde
g_5^2}{(g_5^2+ 2\tilde g_5^2)R \log(R^\prime/R)},
\label{SM2}\\
g^{\prime 2}&=&
\frac{ g_5^2 \tilde g_5^2}{(g_5^2+\tilde g_5^2) R \log(R^\prime/R)},
\label{SM3}
\\
\sin \theta_W &=& \frac{ \tilde g_5}{\sqrt{ g_5^2+2 \tilde g_5^2} }.
\label{SM4}
\eeq
Note, however, that Eqs.~(\ref{tgdef}) and (\ref{edef}) will remain the 
correct definitions of the couplings even to higher orders in the log
expansion. Furthermore, together with Eq.~(\ref{eq:couplings}), they 
determine the remaining normalization factors that are set by requiring 
the correct couplings for the $W$ and $Z$, namely:
\beq
g_5 \psi_1^{(L\pm)} (R) & = & g~,\\
g_5 \psi_1^{(L3)} (R) &=& g \cos \theta_W~.
\eeq
Thus, at the leading order, the normalization coefficients 
in~(\ref{eq:c0Lpm})-(\ref{eq:c1B}) are:

\beq
c_{\pm}&=&\frac{1}{\sqrt{R \log(R^\prime/R)}}, \\
c&=& \sqrt{\frac{g_5^2+\tilde{g}_5^2}{g_5^2+2\tilde{g}_5^2}}
\frac{1}{\sqrt{R \log(R^\prime/R)}},\\
a_0&=& \frac{g_5 \tilde g_5}{\sqrt{(g_5^2+2\tilde{g}_5^2)R 
\log(R^\prime/R)}}.
\eeq

All the oblique corrections  are now contained in the 
wavefunction and mass renormalizations of the gauge
bosons. The wave function renormalizations are given by
\beq
Z_W&=&\int_R^{R'}[\Psi^{W}]^2 \left(\frac{R}{z}\right)
dz=\int_R^{R'}([\Psi^{L+}]^2+[\Psi^{R+}]^2)\left(\frac{R}{z}\right)
dz,\\
Z_Z&=&\int_R^{R'}[\Psi^{Z}]^2\left(\frac{R}{z}\right)
dz=\int_R^{R'}([\Psi^{L3}]^2+[\Psi^{R3}]^2+[\Psi^{B}]^2)\left(\frac{R}{z
}\right)  dz.
\eeq
It is convenient to define equivalent vacuum polarization functions for
these wavefunction renormalizations. Since we have already seen that
$Z_\gamma=1$, the corresponding
vacuum polarization
\beq
\Pi_{QQ}^\prime= \frac{d}{d p^2} \Pi_{QQ}(p^2)|_{p^2=0}
\eeq
vanishes. Also, since we are doing a tree-level calculation,
there is no $Z-\gamma$ mixing and thus $\Pi_{3Q}^\prime$ also vanishes.
We are left with the
simple relations:
\beq
Z_W = 1-g^2 \Pi_{11}^\prime~, \ \ \ \ \ \ \ \
Z_Z = 1-(g^2+g^{\prime\,2}) \Pi_{33}^\prime~.
\eeq
With these definitions we find
\beq
\Pi_{11}^\prime=\Pi_{33}^\prime= \frac{3}{8} \frac{1}{g^2 \log
\frac{R'}{R}}~.
\eeq
Similarly we can calculate the vacuum polarizations at zero momentum, 
corresponding to the mass renormalizations of the gauge bosons:
\beq
g^2 \Pi_{11}(0)&=&\int_R^{R'}([\partial_z
\Psi^{L+}]^2+[\partial_z\Psi^{R+}]^2)\left(\frac{R}{z}\right) dz,\\
(g^2+g^{\prime\,2})\Pi_{33}(0)&=&\int_R^{R'}([\partial_z
\Psi^{L3}]^2+[\partial_z \Psi^{R3}]^2+[\partial_z
\Psi^{B}]^2)\left(\frac{R}{z}\right)  dz.
\eeq
Note that $\Pi_{QQ}(0)$ vanishes because the photon is exactly
massless. We find
\begin{equation}
\Pi_{11}(0)=
\Pi_{33}(0)=\frac{1}{g^2 R'^2 \log \frac{R'}{R}} + \mathcal{O} \left(
\frac{1}{\left(\log \frac{R'}{R}\right)^3}\right)~.
\end{equation}
As a consistency check we can also show that the correction to the W 
mass is correctly
reproduced by the $\Pi's$:
\begin{equation}
M_W^2=g^2 \Pi_{11}(0)(1+g^2 \Pi_{11}')=\frac{1}{R'^2 \log\frac{R'}{R}}
\left(1+\frac{3}{8\log\frac{R'}{R}} \right)~.
\end{equation}
Note, that in the absence of a Higgs VEV the leading contribution to the
W mass itself comes from $\Pi_{11}(0)$, as happens in technicolor 
models as well.

The oblique correction parameters in terms of these vacuum polarization 
functions are defined as~\cite{PT}:
\beq
S&\equiv&16 \pi\, \big(\Pi_{33}^\prime - \Pi_{3Q}^\prime\big)\nonumber 
,\\
T&\equiv&\frac{4 \pi}{\sin^2 \theta_W \cos^2 \theta_W\, M_Z^2}\, 
\big(\Pi_{11}(0) -
\Pi_{33}(0)\big),\\
U&\equiv&16 \pi\, \big(\Pi_{11}^\prime - \Pi_{33}^\prime\big)\nonumber.
\eeq
For fermions localized on the Planck brane we find:
\beq
S&\approx&\frac{ 6 \pi }{g^2\log \frac{R'}{R}}\approx 1.15
\label{eq:S},\\
T&=&0,\\
U&=&0~.
\eeq
Thus just as in technicolor models we find that there is a large 
positive contribution to the $S$ parameter. In technicolor
language this corresponds to the effects of the CFT. The vanishing of 
$T$ is ensured by the custodial SU(2) symmetry close to the
TeV brane. These results are in agreement with the expectations 
presented in~\cite{BPR}.

The experimental values of $S,T,U$ are~\cite{STUexp}
\beq
S&=& -0.03\pm 0.11\\
T&=&-0.02\pm 0.13\\
U&=& 0.24\pm 0.13~.
\eeq
Clearly, with these central values $S$ is too large by a factor of 5-6 
in order to be consistent with the electroweak
precision observables. However, the above experimental values have been 
obtained assuming the existence of a SM Higgs
with mass of 115~GeV. In order to correctly compare this model with the 
experimental result, the contribution of the
Higgs would have to be subtracted. This is however harder than it may 
sound at first, since the Higgs contribution is
finite and gauge invariant at one loop only together with the 
contribution of the W,Z gauge bosons. In this model
what will likely happen is that the Higgs+W,Z contributions have to be 
replaced by the full 1-loop contribution of the
entire W,Z KK towers in order to obtain a gauge invariant answer. This 
would be very interesting to do, here however we will
not follow this route, but rather try to see if the introduction of 
more parameters could reduce the size of the correction to $S$.

\section{Asymmetric gauge couplings and brane localized kinetic
terms.}
\setcounter{equation}{0}
\setcounter{footnote}{0}

In this section, we analyse the more general case where the bulk gauge 
couplings of the $SU(2)_L$ and $SU(2)_R$ are different, $g_{5R}$ and 
$g_{5L}$, and brane kinetic terms allowed by the gauge symmetries of 
the theory are added. Some of these
parameters have already been considered in \cite{Nomura}, where it was 
shown that turning on these parameters does not
change the leading order predictions for the SM masses and couplings. 
The AdS/CFT interpretation of different left-right bulk gauge couplings 
is that the CFT does not have a left-right
interchange symmetry, or in other words is a chiral gauge theory, and 
so in our context such models are analogous to chiral technicolor 
\cite{chiraltc}.
On the Planck brane, the only allowed kinetic terms involve the locally 
unbroken SM group SU(2)$_L \times$U(1)$_Y$.
On the other hand, on the TeV brane the unbroken groups are the 
$U(1)_{B-L}$ and $SU(2)_D$, so that their effect is to mix the SM gauge 
bosons.  In the case of $SU(2)_D$ it is clear that there is a direct 
contribution to the $S$ parameter, not suppressed by a log. In both 
cases there will be corrections to $S$, $T$, and $U$ even when the 
wavefunction is only kept to leading log
order.  For this reason we will
deal with the TeV brane induced terms separately.

\subsection{Planck brane kinetic terms.}

The bulk Lagrangian is as in (\ref{eq:lagr}), except now $g_{5L}\neq 
g_{5R}$.
The localized Lagrangian on the Planck brane is:

\begin{equation} \label{genLagr}
\mathcal{L}_{Pl} = - \left[ \frac{r}{4} {W^L_{\mu \nu}}^2 +
\frac{r'}{4} \frac{1}{g_{5R}^2 + \tilde g_5^2} (g_{5R} B_{\mu \nu} + 
\tilde g_5 W^{R3}_{\mu \nu})^2 \right] \delta (z-R)~,
\end{equation}
where the parameters $r$ and $r'$ have dimension of length.
The presence of the localized kinetic terms modifies the BC's for the 
gauge bosons which now read:
\begin{eqnarray}
&
{\rm at }\  z=R^\prime:
&
\left\{
\begin{array}{l}
\partial_z (g_{5R} A^{L\,a}_\mu + g_{5L} A^{R\,a}_\mu)= 0, \
g_{5L} A^{L\,a}_\mu  - g_{5R} A^{R\,a}_\mu =0, \
\partial_z B_\mu = 0,
\\
\tv{15}
(g_{5R} A^{L\,a}_5 +g_{5L} A^{R\,a}_5) = 0, \
\partial_z (g_{5L} A^{L\,a}_5 - g_{5R} A^{R\,a}_5)  = 0, \
B_5 = 0.
\end{array}
\right.
\label{bc1LR}\\
&{\rm at }\ z=R:
&
\left\{
\begin{array}{l}
\partial_5 A^{L\,a}_\mu + r M^2 A^{L\,a}_\mu=0, \
       A^{R\,1,2}_\mu=0,
\\
\tv{15}
\partial_z (g_{5R}  B_\mu + \tilde g_5  A^{R\,3}_\mu ) + r' M^2 (g_{5R} 
  B_\mu + \tilde g_5  A^{R\,3}_\mu )= 0, \\
\tv{15}
\tilde g_5 B_\mu - g_{5R} A^{R\,3}_\mu=0,
\\
\tv{15}
A^{L\,a}_5=0, \ A^{R\,a}_5=0, \ B_5 = 0.
\end{array}
\right.
\label{bc2LR}
\end{eqnarray}

Following the steps outlined in the previous section, and also
including the effect of the localized Lagrangian in the wave function 
renormalization factors $Z$, the SM
  gauge couplings are defined by
\begin{eqnarray}
\frac{1}{g^2} &=& \frac{R \log \frac{R'}{R} + r}{g^2_{5L}}~,\\
\frac{1}{g'^2} &=& \left( R \log \frac{R'}{R}+r' \right) \left( 
\frac{1}{g^2_{5R}} +
\frac{1}{\tilde{g}_5^2} \right)~.
\end{eqnarray}
With these definitions, all the SM relations are satisfied at leading 
order, so that the oblique corrections are
suppressed by powers of $\log R'/R$.
There are three extra free parameters: the ratio
$g_{5R}/g_{5L}$, $r$ and $r'$.

The BC's in Eq.~(\ref{bc1LR})-(\ref{bc2LR}) also modify the mass 
eigenvalues.
For istance, the equation determining the tower of $W$ masses is:
\begin{multline}
g_{5L}^2 (R_0 - \tilde R_0)(R_1 - \tilde R_1) + g_{5R}^2 (R_1 - \tilde 
R_0)(R_0 - \tilde R_1) +\\ (g_{5L}^2 + g_{5R}^2) (R_1-\tilde R_0)(R_1 - 
\tilde R_1) M r \frac{J_1 (M R)}{J_0 (M R)} = 0\,~,
\end{multline}
where the ratios $R_{0,1}$ and $\tilde R_{0,1}$ are defined by
\beq
R_i = \frac{Y_i (M R)}{J_i (M R)}~, \quad \tilde R_i = \frac{Y_i (M 
R')}{J_i (M R')}~.
\eeq
The lightest solution is:
\beq \label{eq:Wmass}
M_W^2 =
\frac{2}{1 +  \frac{g_{5R}^2}{g_{5L}^2}}
\frac{1}{1+\frac{r}{R \log R'/R}}\,\frac{1}{R'^2 \log\frac{R'}{R}}
\left( 1+\frac{2}{1 + \frac{g_{5R}^2}{g_{5L}^2}}\frac{1}{1+\frac{r}{R 
\log R'/R}}\,
\frac{3}{8\log\frac{R'}{R}} \right)~.
\eeq
Analogously, for the $Z$ mass we find:
\beq \label{eq:Zmass}
M_Z^2 = \frac{2}{1 +
\frac{g_{5R}^2}{g_{5L}^2}}\frac{1}{1+\frac{r}{R \log R'/R}}\,\frac{g^2 
+ g'^2}{g^2} \frac{1}{R'^2 \log\frac{R'}{R}}
\left( 1+\frac{2}{1 + \frac{g_{5R}^2}{g_{5L}^2}}\frac{1}{1+\frac{r}{R 
\log R'/R}}\, \frac{g^2 + g'^2}{g^2}
\frac{3}{8\log\frac{R'}{R}} \right)~.
\eeq

Applying the same formalism as in the previous section for the 
wavefunction and mass renormalizations we find:
\begin{eqnarray}
\Pi_{11}^\prime = \Pi_{33}^\prime &=& \frac{3}{8} \frac{2}{1
+\frac{g_{5R}^2}{g_{5L}^2} } \frac{1}{1+\frac{r}{R \log R'/R}}  
\frac{1}{g^2 \log \frac{R'}{R}}~,\\
\Pi_{11}(0) = \Pi_{33}(0) &=& \frac{2}{1 + \frac{g_{5R}^2}{g_{5L}^2}}  
\frac{1}{1+\frac{r}{R \log R'/R}}
\frac{1}{g^2 R'^2 \log \frac{R'}{R}} + \mathcal{O} \left(
\frac{1}{\left(\log \frac{R'}{R}\right)^3}\right)~.
\end{eqnarray}
As before, we can also reproduce the correct W mass in 
Eq.~(\ref{eq:Wmass}) via the relation:
\begin{equation}
M_W^2=g^2 \Pi_{11}(0)(1+g^2 \Pi_{11}')~.
\end{equation}
Note that since the photon is still exactly massless,
$\Pi_{QQ}(0)=0$ and there is no $Z-\gamma$ mixing.
Moreover, $r'$ only appears in the definition of $g'$ and in 
higher-order terms, so it will not appear in the leading
oblique corrections.

In the localized fermion approximation, we find
\beq
S&\approx&\frac{ 6 \pi }{g^2\log \frac{R'}{R}} \frac{2}{1 +
\frac{g_{5R}^2}{g_{5L}^2}} \frac{1}{1+\frac{r}{R \log R'/R}} 
,\label{eq:SPl}\\
T& \approx &0,\\
U&\approx&0~.
\eeq
Note that both the leading contribution to the $S$ parameter and the 
$W$ mass square are multiplied by the same suppression factor.
This means that if we want to lower the numerical value of $S$ in 
Eq.~(\ref{eq:S})
by a factor of $10$, the value of $R'$ decreases by a factor of 
$1/\sqrt{10}$.
Indeed, we have numerically checked that the mass of the first $W$ 
resonance is increased
from $\sim 1200$~GeV to $\sim 3800$~GeV. This means that increasing the 
ratio $g_{5R}/g_{5L}$ or $r$ is not an
effective way of reducing the contribution to the $S$ parameter: as $S$ 
gets reduced, the KK modes of W and Z get pushed
to higher values, and thus perturbative unitarity will be lost when 
$M_{W'},M_{Z'}$ get pushed above 1800~GeV. After that
a tree-level calculation is no longer reliable, therefore a  claim that 
for such parameters $S$ is significantly
reduced is probably also not reliable.

\subsection{TeV brane kinetic terms: linear analysis}

The localized Lagrangian on the TeV brane is
\begin{equation} \label{genLagrTeV}
\mathcal{L}_{brane} = - \frac{R'}{R} \left[ \frac{\tau'}{4} B_{\mu 
\nu}^2
+ \frac{\tau}{4} \frac{1}{g_{5R}^2 + g_{5L}^2} (g_{5R} W^L_{\mu \nu} + 
g_{5L} W^{R}_{\mu \nu})^2 \right] \delta (z-R')~,
\end{equation}
where the parameters $\tau$ and $\tau'$ have dimension of length, and 
the modified BC for the gauge bosons are:
\begin{eqnarray}
&
{\rm at }\  z=R^\prime:
&
\left\{
\begin{array}{l}
\partial_z (g_{5R} A^{L\,a}_\mu + g_{5L} A^{R\,a}_\mu) - \tau M^2  
\frac{R'}{R} (g_{5R} A^{L\,a}_\mu + g_{5L} A^{R\,a}_\mu) = 0, \\
\tv{15}
g_{5L} A^{L\,a}_\mu  - g_{5R} A^{R\,a}_\mu =0, \
\partial_z B_\mu - \tau' M^2 \frac{R'}{R} B_\mu  = 0,
\\
\tv{15}
(g_{5R} A^{L\,a}_5 +g_{5L} A^{R\,a}_5) = 0, \
\partial_z (g_{5L} A^{L\,a}_5 - g_{5R} A^{R\,a}_5)  = 0, \
B_5 = 0.
\end{array}
\right.
\label{bc1TeV}\\
&{\rm at }\ z=R:
&
\left\{
\begin{array}{l}
\partial_5 A^{L\,a}_\mu + r M^2 A^{L\,a}_\mu=0, \
       A^{R\,1,2}_\mu=0,
\\
\tv{15}
\partial_z (g_{5R}  B_\mu + \tilde g_5  A^{R\,3}_\mu ) + r' M^2 (g_{5R} 
  B_\mu + \tilde g_5  A^{R\,3}_\mu )= 0, \\
\tv{15}
\tilde g_5 B_\mu - g_{5R} A^{R\,3}_\mu=0,
\\
\tv{15}
A^{L\,a}_5=0, \ A^{R\,a}_5=0, \ B_5 = 0.
\end{array}
\right.
\label{bc2TeV}
\end{eqnarray}

As already mentioned, such kinetic terms are not diagonal in the SM 
gauge group, so they can make a large contribution to the $S$ 
parameter.
We therefore expect a contribution unsuppressed by  $\log R'/R$, that 
can in principle cancel the contribution found in the previous 
subsection. In order to identify the leading contribution, we only need 
to expand the wave functions to order 1/log.
We will then expand the results for small kinetic terms, allowing also 
to simplify the analytical formulae.

At leading order in the log, the SM gauge couplings are set by the 
following relations, which follow from Eqs.~(\ref{tgdef}) and
(\ref{edef}) (where the localized terms are taken into account in the 
normalization):
\beq
\frac{1}{e^2} & = & \frac{R \log R'/R + r + \tau}{g_{5L}^2} + \frac{R 
\log R'/R + r' + \tau}{g_{5R}^2} + \frac{R \log R'/R + r' + 
\tau'}{\tilde g_{5}^2}~,\\
\frac{g'^2}{g^2} &=& - \frac{\tilde g_5 \, \psi_Z^{(B)} (R)}{g_{5L} \, 
\psi_Z^{(L3)} (R)}~,
\eeq
where the latter is completely determined by the BC's (independent of 
the normalization of the wave functions).

Linearizing for small $\tau/(R \log R'/R) \ll 1$ and $\tau'/(R \log 
R'/R) \ll 1$, it follows
\begin{eqnarray}
\frac{1}{g^2} &=& \frac{R \log R'/R + r+ 
\tau}{g_{5L}^2}~,\label{eq:gTeV}\\
\frac{1}{g'^2} &=& \frac{R \log R'/R + r'+ \tau}{g_{5R}^2}+\frac{R \log 
R'/R + r'+ \tau'}{\tilde g_{5}^2}~. \label{eq:gpTeV}
\end{eqnarray}

The $W$ and $Z$ masses are also corrected by
\beq \label{eq:WZmassTeV}
M_W^2 &=& M_{W0}^2
\left( 1 - \frac{g_{5R}^2}{g_{5R}^2+g_{5L}^2} \frac{\tau}{r+R \log 
R'/R} \right)~, \\
M_Z^2 &=& M_{Z0}^2 \left( 1- \frac{g_{5R}^2}{g_{5R}^2 + 
g_{5L}^2}\frac{\tau}{r+R \log R'/R} \left(1- \frac{g_{5L}^2 
g'^2}{g_{5R}^2 g^2}\right) \right)~,
\eeq
where $M_{W0}$ and $M_{Z0}$ are defined in 
Eqs.~(\ref{eq:Wmass})-(\ref{eq:Zmass}), in terms of the corrected $g$ 
and $g'$,
given in Eqs.~(\ref{eq:gTeV})-(\ref{eq:gpTeV}).

Applying the formalism from the previous section,  we find the vacuum 
polarization functions at leading log :
\begin{eqnarray}
\Pi_{11}^\prime = \Pi_{33}^\prime &=& \frac{1}{1
+\frac{g_{5R}^2}{g_{5L}^2} } \frac{\tau}{r + R \log R'/R}  
\frac{1}{g^2}~,
\\
\Pi_{11}(0) = \Pi_{33}(0) &=& \frac{2}{1 + \frac{g_{5R}^2}{g_{5L}^2}}  
\frac{1}{1+\frac{r}{R \log R'/R}}
\frac{1}{g^2 R'^2 \log \frac{R'}{R}}
\left( 1 - \frac{\tau}{r + R \log R'/R}  \right)
~.
\end{eqnarray}
As before, we can also reproduce the correct W mass in 
Eq.~(\ref{eq:Wmass}) via the relation:
\begin{equation}
M_W^2=g^2 \Pi_{11}(0)(1+g^2 \Pi_{11}')~,
\end{equation}
and analogously for the $Z$ mass. Note that again the photon is exactly 
massless
($\Pi_{QQ}(0)=0$) and there is no $Z-\gamma$ mixing.

In the localized fermion approximation, we find
\beq
\delta S&\approx&\frac{ 8 \pi }{g^2}\frac{2}{1 +
\frac{g_{5R}^2}{g_{5L}^2}} \frac{\tau}{r+R \log R'/R}~, 
\label{eq:Stau}\\
\delta  T& \approx &0~,\\
\delta  U&\approx&0~.
\eeq

Note that in order to obtain the full expression for $S$, (\ref{eq:SPl})
has to be added to the result above:
  \beq
S\approx\frac{ 6 \pi }{g^2 \log R'/R}\frac{2}{1 +
\frac{g_{5R}^2}{g_{5L}^2}} \frac{1}{1+\frac{r}{R \log R'/R}} \left( 
1+\frac{4}{3} \frac{\tau}{R} \right)~.
\eeq
This means that $S$ would vanish if $\tau \sim - 3/4 R$.
In this case the expansion parameter used in the computation is still 
small, being divided by a log.
However, the wrong sign kinetic term introduces a tachyonic solution in 
the spectrum of the $W$'s.
For small values of $\tau$, the mass can be large and well above the 
cutoff of the theory, but we have checked that in order to cancel $S$ a 
tachyon at around $800$~GeV appears.

\subsection{TeV brane kinetic terms: beyond linear perturbation for the 
$B-L$ kinetic term}

As we have seen in the previous subsection, the $B-L$ kinetic term 
$\tau'$ does not correct the oblique parameters at linear level. 
However, as we are going to show now, at the non-linear level it 
provides a negative contribution to the $S$ parameter. We are going to 
show this result by working analytically
to quadratic order in an expansion in $\tau'/(R  \log R^\prime/R)$. We 
will also give numerical
results beyond this quadratic level to emphasize that the negative 
contribution to $S$ is not an expansion artifact.  Since our point here 
is not to give a general analysis of the parameter space but simply to
exhibit a way to suppress $S$, we analyse the simple case where the 
other localized terms are negligible and $g_{5L} = g_{5R} = g_5$.
It is straightforward to follow the same formalism, so that at leading 
log  and quadratic order in $\tau'$ we find:
\begin{eqnarray}
\Pi_{11}^\prime = \Pi_{33}^\prime &=& -\frac{1}{2(g_{5}^2 + \tilde 
g_{5}^2) } \frac{\tau'^2}{R \log R'/R}~,\\
\Pi_{11}(0) &=& \frac{R}{g_{5}^2 R'^2}~, \\
\Pi_{33}(0) &=& \frac{R}{g_{5}^2 R'^2} \left( 1+\frac{g_{5}^2 \tilde 
g_{5}^2}{2 (g_{5}^2 + \tilde g_{5}^2)^2} \frac{\tau'^2}{(R \log 
\frac{R'}{R})^2} \right)~.
\end{eqnarray}

In this simple case, the only free parameter that we have at our 
disposal is the value of the localized kinetic term, $\tau^\prime$. The 
values of the 5D gauge couplings, $g_5$ and ${\tilde g}_5$ are related
to the SM gauge couplings, at the quadratic order in $\tau'$, by
\begin{eqnarray}
\frac{1}{g^2} + \frac{1}{g'^2} =
\frac{1}{g_5^2} \left( 2 + \left( 1+\frac{\tau'}{R \log R'/R}\right) 
\frac{g_5^2}{{\tilde g}_5^2} \right) R \log \frac{R'}{R}
~,\\
\frac{g_5^2}{{\tilde g}_5^2} =  \frac{g^2-g'^2}{g'^2}
\left( 1 - \frac{\tau'}{R \log R'/R} + \frac{g^2-g'^2}{2g^2} 
\frac{\tau'^2}{R^2 \log^2 R'/R} \right).
\end{eqnarray}

\begin{figure}[tb]
\begin{center}
\includegraphics[width=12cm]{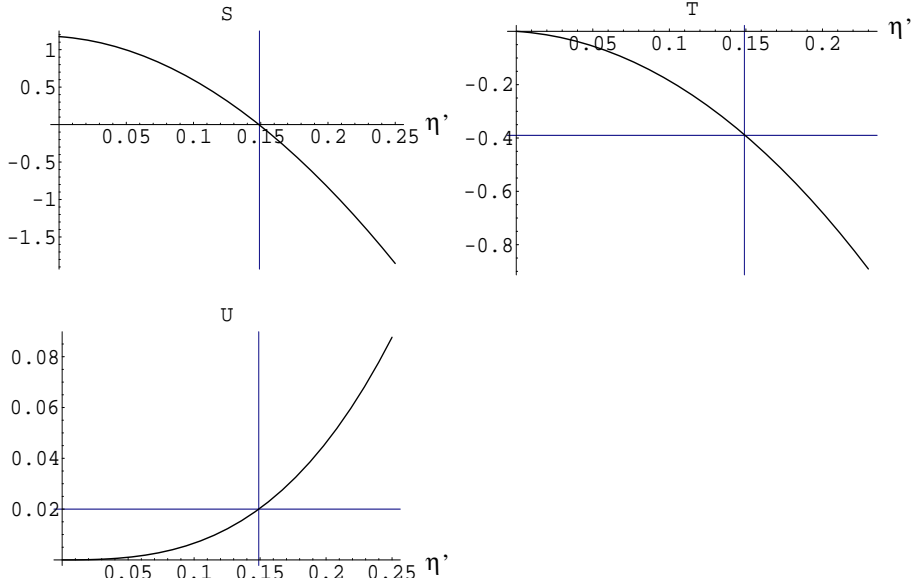}
\end{center}
\caption{Plot of the oblique parameters as functions of the $B-L$ 
kinetic term on the TeV brane,
normalized as $\eta' = \tau' /(R \log R'/R)$. The grid lines point to 
the region where $S$ vanishes. 
 These results include higher order corrections in the log and $\tau'$ expansion, and we also checked them numerically.} \label{fig:STU}
\end{figure}

The expressions  obtained for $S,T,U$ are:
\beq
S&\approx&\frac{ 6 \pi }{g^2\log \frac{R'}{R}}-\frac{ 8 \pi }{g^2} 
\left( 1-\left(\frac{g'}{g}\right)^2 \right) \frac{\tau'^2}{(R \log 
R'/R)^2}~, \label{eq:Staup}\\
T&\approx&-\frac{ 2 \pi }{g^2} \left( 1-\left(\frac{g'}{g}\right)^4 
\right) \frac{\tau'^2}{(R \log R'/R)^2}~,\\
U&\approx&0~.
\eeq

In this case, as $g'/g = \tan \theta_W < 1$, the additional 
contribution to $S$ is negative.
This means that this contribution can cancel the positive 
log-suppressed result in Eq.~(\ref{eq:S}).
In Fig.~\ref{fig:STU} we plotted the numerical values of the oblique 
parameters, as functions of the properly normalized $B-L$ kinetic term.
$S$ vanishes for $\tau'/(R \log R'/R) \approx 0.15$, a small value.
We also find a negative contribution to $T$, that is $\approx -0.4$ 
when $S$ vanishes
and a small contribution to $U$ from higher orders, $U \approx 0.02$.
However, to T there will also be other moderate positive contributions 
of order $\sim 0.1$ at the
loop level from the $t',b'$ sector.

\begin{figure}[tb]
\begin{center}
\includegraphics[width=10cm]{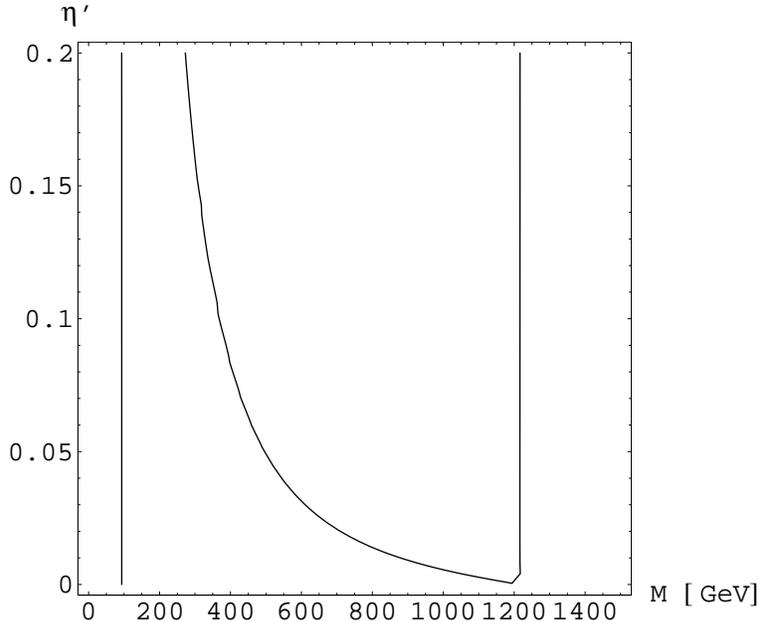}
\end{center}
\caption{The spectrum of the first three $Z$ resonances as a function 
of the $B-L$ kinetic term, normalized as $\eta' = \tau' /(R \log 
R'/R)$.} \label{fig:spectrum}
\end{figure}

It is now important to examine how the spectrum is changed.
The $W$ resonances are clearly left unchanged, while in 
Fig.~\ref{fig:spectrum} we plotted the masses of the first three $Z$ 
resonances.
The $Z$ mass is clearly left unchanged, being an imput parameter.
However, the degeneracy of the first resonances originally at 1200~GeV 
is split.
One boson is left stable around $1200$~GeV, while the other becomes 
light and if $S=0$ its mass drops to $\approx 300$~GeV. Increasing the 
$g_{5R}/g_{5L}$ mass ratio one can raise the Z$'$ mass to about 400~GeV 
before perturbative
unitarity is lost.
In other words, this scenario predicts a light Z$'$.
Its couplings to the fermions are fixed by the wave function on the 
Planck brane:
\beq
g_5 \psi_{Z'}^{(L3)}(R)\, T_3 + \tilde g_5 \psi_{Z'}^{(B)}(R)\, 
\frac{Y}{2}~.
\eeq
We have numerically computed:
\beq
\frac{\psi_{Z'}^{(L3)}(R)}{\psi_{Z}^{(L3)}(R)}\approx 0.03\,~, \qquad  
\frac{\psi_{Z'}^{(B)}(R)}{\psi_{Z}^{(B)}(R)}\approx 0.8\,~,
\eeq
so that our extra Z$'$ mainly couples as a sequential hypercharge Z$'$.
Numerically, compared with the electric charge, the coupling is 
$\approx 0.2\, e\, Y/2$.
This particular case is not fully covered in the literature \cite{App}, 
however the coupling should
  be small enough to escape direct production limits, mainly coming from 
the Tevatron (and also at LEP2).
Another issue is the presence of induced 4-fermion couplings. Bounds on 
Z$'$ masses vary between 150 and 800~GeV from
4 Fermi operators at HERA~\cite{HERA}, depending on the precise 
structure and strength of the coupling of the extra Z$'$ to
SM fermions. However, the particular case of a sequential hypercharge 
Z$'$ needs to be investigated in detail, which is beyond the scope
of this paper.

\section{Conclusions}
\setcounter{equation}{0}
\setcounter{footnote}{0}

We have calculated the tree-level oblique corrections to electroweak 
precision observables generated in
higgless models of
electroweak symmetry on a warped background. In the absence of brane 
induced kinetic terms (and equal left and right gauge couplings) we 
find the
$S$ parameter to be $\sim 1.15$, while $T,U\sim 0$. Planck brane 
induced kinetic terms
and unequal left-right couplings can lower $S$, however for 
sufficiently low values of $S$ tree-level unitarity
will be lost. A kinetic term localized on the TeV brane for SU(2)$_D$ 
will generically increase $S$, however
an induced kinetic term for U(1)$_{B-L}$ on the TeV brane will lower 
$S$. With appropriate choice of the value of this
induced kinetic term $S\sim 0$ can be achieved. In this case the mass 
of the lowest Z$'$ mode will be lowered to $\sim
300$~GeV.

\section*{Acknowledgments}

While this paper was in preparation, two related studies~\cite{DHLR,BN}
 of the oblique corrections to electroweak precision observables in 
higgsless models appeared, which have a
considerable overlap with our work.

We thank Maxim Perelstein, Riccardo Rattazzi, Neal Weiner and Jim Wells
for useful discussions and comments.
The research of G.C. and C.C.
is supported in part by the DOE OJI grant DE-FG02-01ER41206 and in part
by the NSF grants PHY-0139738  and PHY-0098631.
C.G. is supported in part by the RTN European Program
HPRN-CT-2000-00148, by the ACI Jeunes Chercheurs 2068
and by the Michigan Center for Theoretical Physics. J.T. is supported 
by the US Department of Energy under contract
W-7405-ENG-36.


\end{document}